\documentclass[a4paper,11pt]{article}
\usepackage{mathtools}
\usepackage{amsmath, amsthm, amssymb}
\usepackage{mathrsfs}
\usepackage{jcappub} 
\usepackage{lineno}

\arxivnumber{1234.56789} 
\title{\textit{Tele-Correlation}: Calibrating Shear-Shear Correlation with Real Data}

\author[a]{Zhi Shen}
\author[a,b]{Jun Zhang}
\author[a]{Cong Liu}
\author[c]{Hekun Li}
\author[a]{Haoran Wang}
\author[a]{Zhenjie Liu}
\author[a]{Jiarui Sun}

\affiliation[a]{Department of Astronomy, Shanghai Jiao Tong University, Shanghai 200240, China}
\affiliation[b]{Shanghai Key Laboratory for Particle Physics and Cosmology, Shanghai 200240, China}
\affiliation[c]{Shanghai Astronomical Observatory, Chinese Academy of Sciences, Shanghai 200030, China}
\emailAdd{betajzhang@sjtu.edu.cn}

\abstract{Tele-correlation refers to the correlation of galaxy shapes with large angular separations (e.g., $>100$ degrees). Since there are no astrophysical reasons causing such a correlation on cosmological scales, any detected tele-correlation could disclose systematic effects in shear-shear correlation measurement. If the shear estimators are measured on single exposures, we show that the field distortion (FD) signal associated with the galaxy position on the CCD can be retained and used in tele-correlation to help us directly calibrate the multiplicative and additive biases in shear-shear correlations. 
We use the DECaLS shear catalog produced by the Fourier\_Quad pipeline to demonstrate this idea.
To our surprise, we find that significant multiplicative biases can arise (up to more than 10\%) due to redshift binning of the galaxies. Correction for this bias leads to about 1$\sigma$ increase of the best-fit value of $S_8$ from $0.760^{+0.015}_{-0.017}$ to $0.777^{+0.016}_{-0.019}$ in our tomography study. }

\begin{document}
\maketitle
\flushbottom

\section{Introduction}
\label{sec:intro}
Weak lensing has been established as a powerful probe of the cosmic structure. Many large scale galaxy surveys set weak lensing as their primary scientific goals, including Canada-France-Hawaii Telescope Lensing Survey (CFHTLenS\footnote{\href{http://www.cfhtlens.org/}{www.cfhtlens.org/}})\citep{ref:heymans2013}, Dark Energy Survey (DES\footnote{\href{https://www.darkenergysurvey.org/}{www.darkenergysurvey.org}}) \citep{ref:des_y3}, Hyper Suprime-Cam
Subaru Strategic Program (HSC\footnote{\href{https://hsc.mtk.nao.ac.jp/ssp/}{hsc.mtk.nao.ac.jp/ssp/}})\citep{ref:Lixc2023}, Kilo-Degree Survey (KiDS\footnote{\href{https://kids.strw.leidenuniv.nl/}{kids.strw.leidenuniv.nl}})\citep{ref:asgari2021}. In these studies, it has been found that the parameter $S_8$ from weak lensing is smaller than that from the cosmic microwave background measurement \citep{ref:planck_result} at about $2-3 \sigma$ level. To firmly establish this result, however, we need to carefully examine all possible sources of systematic errors, which is well known to be difficult. We expect these issues to be better addressed in Stage IV surveys, including Euclid \citep{ref:Laureijs2011},  the Large Synoptic Survey Telescope\footnote{\href{https://www.lsst.org/}{www.lsst.org/}}(LSST), the China Space Station Telescope (CSST) \citep{ref:Gong2019}, and Roman \citep{ref:roman,ref:Yamamoto2022}, all of which are going to observe billions of galaxy images for accurate weak lensing measurement. 

So far, major weak lensing surveys calibrate the systematics using image simulations. For example, CFHTLenS takes image simulations to quantify the dependence of shear bias on the signal-to-noise-ratio (SNR) and galaxy size \citep{ref:Heymans2012,ref:Miller2013}; HSC calibrates the multiplicative and additive bias as a function of SNR and image resolution in simulations \citep{ref:Lixc2022}; KiDS team calibrates the shear catalog with a comprehensive model considering the SNR, the galaxy size and ellipticity, as well as the size and ellipticity of the point spread function (PSF) \citep{ref:Kannawadi2019}; The DES team corrects the multiplicative bias with a multiple effective redshift distribution, which is achieved using simulations \citep{ref:Gatti2021,ref:MacCrann2022}.

Ideally, one would hope to test the shear recovery accuracy directly on the real data. One way to do so is to make use of the field distortion (FD) signal that naturally exists in all optical images, as proposed in \cite{ref:jzhang2019}. By grouping the galaxies according to their underlying FD signals, and observing how well the galaxy shear estimators can recover the FD (measured in astrometry), one can get a direct estimate of the multiplicative and additive shear biases. This idea has been successfully applied to the processing of the CFHTLenS data \citep{ref:jzhang2019} and the DECaLS data \citep{ref:jzhang2022} with the Fourier\_Quad shear measurement method, and has helped discover a selection effect due to the presence of image boundaries \citep{Wang_2021}. This kind of test automatically involves all observational and instrumental effects, and has been proved as a robust way for calibrating shape measurement. 

In this work, we show that the FD test can be extended to calibrate potential biases for the two-point statistics, i.e., shear-shear correlations, which is commonly believed to be more challenging to measure due to its small amplitude ($\lesssim 10^{-4}$). We note that this is not completely equivalent to the one-point test, because even if the underlying shear error is close to zero on average, it may still have non-zero spatial correlations due to residual systematic effects on the focal plane. 

The idea of the two-point (2pt hereafter) FD test is to cross-correlate the shear estimators of two remotely separated galaxies, which we call \textit{Tele-Correlation} (TC hereafter). Such a correlation should not contain contributions from the astrophysical signals. If the galaxy shear estimator is measured on each exposure individually, so that the FD shear information is kept in the catalog for each galaxy image, the galaxy pairs involved in the TC measurement can be grouped according to the products of their underlying FD shears, forming a straightforward way of calibrating for the potential biases in the shear-shear correlation.

In this paper, we apply the \textit{Tele-Correlation} method on the DECaLS shear catalog produced by the Fourier\_Quad (FQ hereafter) shear measurement method \citep{ref:jzhang2022}. In \S\ref{sec:dnr} we introduce the DECaLS shear catalog, and the methods we use for the shear estimation and shear-shear correlation. In \S\ref{sec:tc_cssc} we introduce the concept of TC, and show the results from TC, mainly including the shear biases from each redshift bin for our tomographic study. We show how the biases change our cosmology constraints in \S\ref{cosmo}. Finally, we give our main conclusion in \S\ref{sec:conclu_disc}. 

\section{Data and Shear Catalog}\label{sec:dnr}

Our shear catalog is based on the imaging data of the Dark Energy Camera Legacy Survey (DECaLS)\footnote{\href{https://www.legacysurvey.org}{www.legacysurvey.org}}. The total sky coverage is about 10,000 $\mathrm{deg}^2$, taken by the Dark Energy Camera (DECam) on the Blanco 4 m telescope of the Cerro Tololo Inter-American Observatory. The image files are pre-processed through the “Community Pipeline” to remove the instrumental effects, and the sky backgrounds are kept. The shear catalog is obtained using the Fourier\_Quad pipeline, which evolves from the FQ shear measurement method \citep{ref:jzhang2008,ref:jzhang2011,ref:jzhang2011_1,ref:jzhang2015,ref:jzhang2017,ref:jzhang2019,ref:jzhang2022}. It involves all the necessary steps for achieving accurate galaxy shape measurement, including background removal, astrometric calibration, PSF reconstruction, etc.. The pipeline is applied to the data of all three bands: g, r, z. With the field distortion test, it is found that the quality of the z-band shear catalog is much better than those of the other two bands. We therefore only use the z-band data in this work for the shear-shear correlation. The typical galaxy number density is about 3–5 per square arcmin \citep{ref:jzhang2022}. 
The photo-$z$ we use in this paper is measured by \cite{ref:zhou2021}, and for cross-checking the redshift, we also use the photo-$z$ catalog from \cite{ref:Zou2019}. In the following tests, we set a sample cut with those criteria: we choose those galaxies from the z-band with signal-to-noise ratio $\nu_F$ (defined in Fourier space, see \cite{ref:hkli2021}) is larger than 4. Additionally, we exclude galaxies with large field distortion signals: $|g^{f}_{1,2}|> 0.0015$, which are located near the edge of the exposure. We also exclude galaxies obtained from the problematic chips and those with the z-band magnitude larger than 21. After these cuts, there are about one galaxy per square arcmin. Given that the FQ shear estimators are measured on individual exposures, we only use the shear estimators from two different exposures to measure their correlations. This is for the purpose of avoiding possible systematic errors from correlated PSF residuals \citep{ref:Lu2017}.

The FQ shear measurement method uses the multipole moments of the galaxy power spectrum to form the shear estimators. They are defined as:
\begin{equation}
\label{FQ_shear_estimator}
\begin{aligned}
    G_1&=-\frac{1}{2}\int d^2 \vec{k}(k_x^2-k_y^2)T(\vec{k})M(\vec{k})\\  
    G_2&=-\int d^2\vec{k}k_x k_y T(\vec{k})M(\vec{k})\\
    N&=\int d^2 \vec{k} [k^2-\frac{\beta^2}{2}k^4 ]T(\vec{k})M(\vec{k})\\
    U & =-\frac{\beta^2}{2} \int d^2 k(k_x^4-6 k_x^2 k_y^2+k_y^4) T(\vec{k}) M(\vec{k})\\
    V&=-2 \beta^2 \int d^2 k(k_x^3 k_y-k_x k_y^3) T(\vec{k}) M(\vec{k}),
\end{aligned}
\end{equation}
in which $M(\vec{k})$ is the 2D galaxy power spectrum corrected by terms related to the background noise and the Poisson noise \citep{ref:jzhang2015}. $T(\vec{k})$ is the factor for converting the PSF to a Gaussian form, i.e.:   
\begin{equation}
T(\vec{k})={\big|\widetilde{W}_{\beta}(\vec{k})\big|}^2/{\big|\widetilde{W}_{PSF}(\vec{k})\big|}^2\label{tk}
\end{equation}
in which $\widetilde{W}_{PSF}(\vec{k})$ and $\widetilde{W}_{\beta}(\vec{k})$ $[=\exp(-\beta^2\big|\vec{k}\big|^2/2)]$ are the Fourier transforms of the PSF function and the Gaussian kernel respectively. $\beta$ is the scale radius of the kernel, which is chosen to be slightly larger than that of the original PSF, so that the reconvolution defined by $T(\vec{k})$ is mathematically well defined. It can be shown that the FQ shear estimators defined in eq.(\ref{FQ_shear_estimator}) can recover the underlying shear signals to the second order in accuracy:
\begin{equation}
\label{stat1}
\frac{<G_1>}{<N>}=g_1+\mathcal{O}(g_{1,2}^3),\;\frac{<G_2>}{<N>}=g_2+\mathcal{O}(g_{1,2}^3)
\end{equation}

Although the form defined in eq.(\ref{stat1}) is unbiased, it is not statistically optimal, as the amplitudes of the $G_i$ and $N$ are unnormalized moments and therefore strongly depend on the galaxy luminosity. Certain weightings can be applied to reduce the scatter of the shear estimators, but the weighting function should be carefully chosen to avoid additional biases \citep{ref:hkli2021}. Alternatively, in this work, we adopt the PDF-SYM method proposed by \cite{ref:jzhang2017} to measure the stacked shear signal and the shear-shear correlation. In this new method, the shear signal is recovered by symmetrizing the probability distribution function (PDF) of the FQ shear estimator $\hat{G_i}$ with an assumed shear value $\hat{g_i}$ through $\hat{G_i}=G_i-\hat{g_i}B_i$, in which $B_1 =N+U$, $B_2 = N-U$. Note that the term $U$ here and $V$ defined in eq.(\ref{FQ_shear_estimator}) are the two components of a spin-4 quantity, therefore both of them should be kept in the catalog. Similarly, to measure the shear-shear correlation, the joint PDF of two shear estimators $(\hat{G_i},\hat{G_i}^{\prime})$ is symmetrized using an assumed correlation strength. With this method, we can bring the statistical error down to the lower bound without introducing systematic errors. The details of the PDF-SYM method can be found in \cite{ref:jzhang2017}. Its performance has been proved in a number of recent works \citep{ref:mfong2022,ref:jqwang2022,ref:zwzhang2022,ref:hrwang2023}. Its application on the measurement of the shear-shear correlation is also discussed specifically in our another recent work \citep{ref:zjliu2023}.

\section{Tele-Correlation}\label{sec:tc_cssc}

For simplicity and clarity, we assume that the shear estimator takes a conventional form $e_i$ with $i=1,2$, which can be regarded as the two galaxy ellipticity components. Let us also denote the cosmic shear components and the convergence as $\gamma_i$ and $\kappa$. The reduced shear $g_i$ is therefore given by $\gamma_i/(1-\kappa)$. Under the weak lensing effect and the field distortion effect, the shear estimator would take the following form: 
\begin{equation}
    e_i = e^I_{i}+ (1+m_i)(g_i + g^f_i)+n_i
\end{equation}
where $e^I_i$ is the intrinsic ellipticity, and $g^f_i$ is the field distortion signal. $m_i$ is the multiplicative bias associated with the shear estimator, and $n_i$ is the noise, possibly containing an additive bias. 
\textit{Tele-Correlation}(TC) is defined as the correlation between the shear estimators of two galaxies separated by a large distance (e.g., more than a hundred degrees). In this case, the astrophysical correlation should vanish, and TC only receives contribution from the field distortion:
\begin{eqnarray}
\label{ee_infty}
&&<e_ie_i^{\prime}>(\Delta\theta\rightarrow\infty)\\ \nonumber
&=&(1+m_i)(1+m_i')<g_i^{f}g_i^{f\prime}>+<n_in_i'>\\ \nonumber  
&=&(1+\mathcal{M}_i)^2<g_i^{f}g_i^{f\prime}>+\mathcal{C}_i
\end{eqnarray}
We denote TC as $<e_ie_i^{\prime}>(\Delta\theta\rightarrow\infty)$ in the rest of the paper. It is clear that by grouping the galaxy pairs according to the products of their FD signals, TC can be directly compared with $<g_i^{f}g_i^{f\prime}>$ to estimate $\mathcal{M}_i$ and $\mathcal{C}_i$, and thereby the multiplicative and additive biases of the shear estimators. This calibration can be done for galaxies of different redshift bins. For the auto-correlation between the same population of the galaxies, we should have $\mathcal{M}_i=m_i$. 

\begin{figure}
\centering
\includegraphics[scale=0.7]{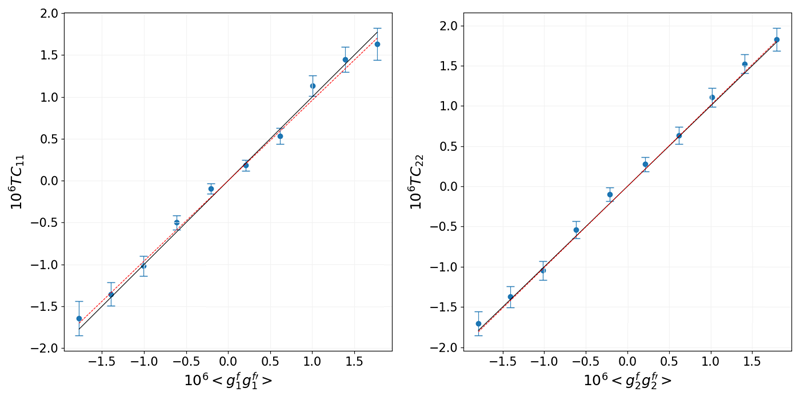}
\caption{TC of galaxy pairs with angular distance larger than 140 degrees. In each panel, the black solid line is 'y=x', and the red dashed line is the best-fit line of data points. The $TC_{11}$ and $TC_{22}$ parts are shown in the left and right panels respectively. The multiplicative and additive biases for the left and right panels are $\mathcal{M}_1=-(1.99\pm 1.72)\times 10^{-2}$, $\mathcal{C}_1=(4.1\pm2.6)\times 10^{-8}$, $\mathcal{M}_2=(0.48\pm 0.78)\times 10^{-2}$, and $\mathcal{C}_2=(6.4\pm1.53)\times 10^{-8}$ respectively.}
\label{fig:corr_fd}
\end{figure}

Fig.\ref{fig:corr_fd} shows the result of TC using the galaxy pairs separated by more than 140 degrees. We only choose the galaxies with redshifts larger than 0.1. The FD product (x-axis in the figure) is evenly divided into 10 bins. The red dashed lines are the best-fit of the data points, and the black solid lines refer to 'y=x'. According to the figure, the multiplicative biases are not significant: $m_1 = (-1.99\pm1.717)\times 10^{-2}$; $m_2=(0.48\pm0.779)\times 10^{-2}$, and the additive biases $\mathcal{C}_i$ are negligible $(\sim 10^{-8})$. 

It is interesting to note that one may expect to get smaller error bars in TC if we enlarge the angular range of the galaxy pairs to get more samples. 
This is indeed the case: we find that the average error-bar size of the data points in fig.\ref{fig:corr_fd} does decay when the lower limit of the separation angle is reduced. However, the values and the uncertainties of the multiplicative biases do not change much. In practice, when the lower limit of the galaxy angular separation is set at 140 degree, the multiplicative biases converge well enough, and the computational cost is already quite substantial\footnote{More than 70000 cores$\cdot$h for computing $1.7*10^{15}$ galaxies pairs.}. As a consistency check, we also set the angular range of the galaxy pairs to be between 120 to 140 degrees, and the results are similar to those shown in fig.\ref{fig:corr_fd}.
\begin{figure}
\centering
\includegraphics[scale=0.7]{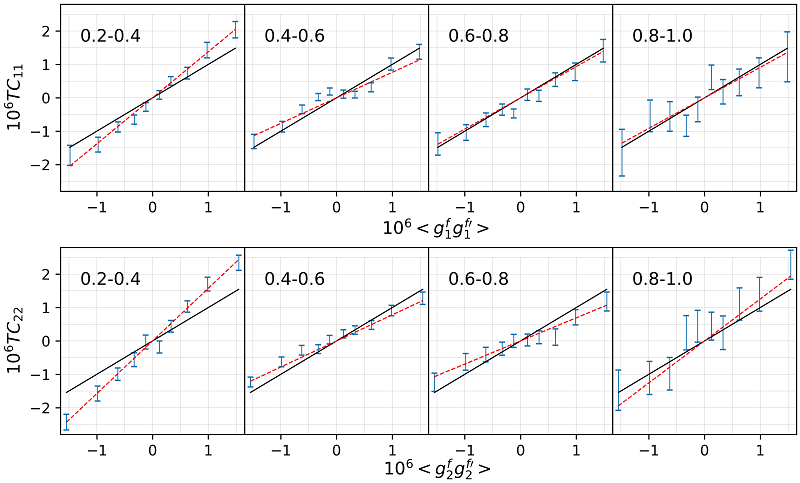}
\caption{The TC of galaxies in different redshift bins using the photo-z from \cite{ref:zhou2021}. $TC_{11}$ and $TC_{22}$ are shown in the upper and lower rows. In each panel, the redshift range is shown in the upper-left corner. The red dashed line is the best-fit for the data points, and the black solid line is the 'y=x' line. 
The multiplicative and additive biases are shown in table.\ref{tab:mc_of_db}.}
\label{fig:2Dfd_db}
\end{figure}

\begin{figure}
    \centering
    \includegraphics[scale = 0.7]{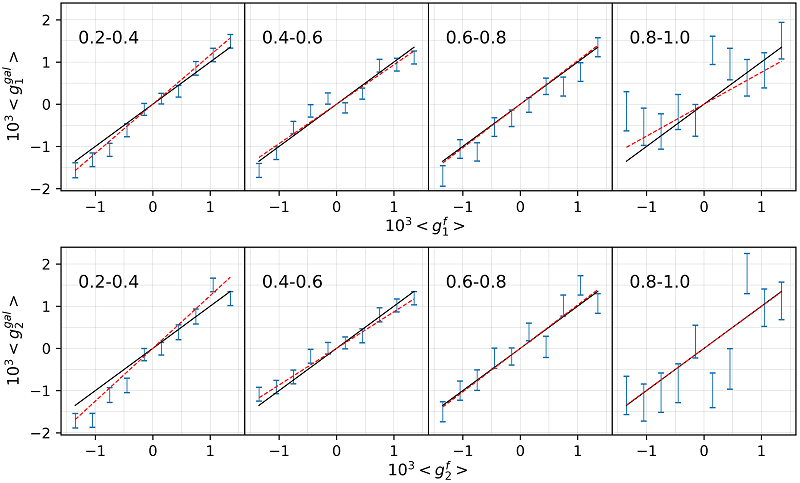}
    \caption{The results of the original FD test (shear stacking) for different redshift bins. In each panel, the redshift range is indicated in the upper-left corner, and the red dashed line is the best-fit of the data points.}
    \label{fig:1Dfd_db}
\end{figure}

\begin{figure*}
   \centering
    \includegraphics[scale = 0.7]{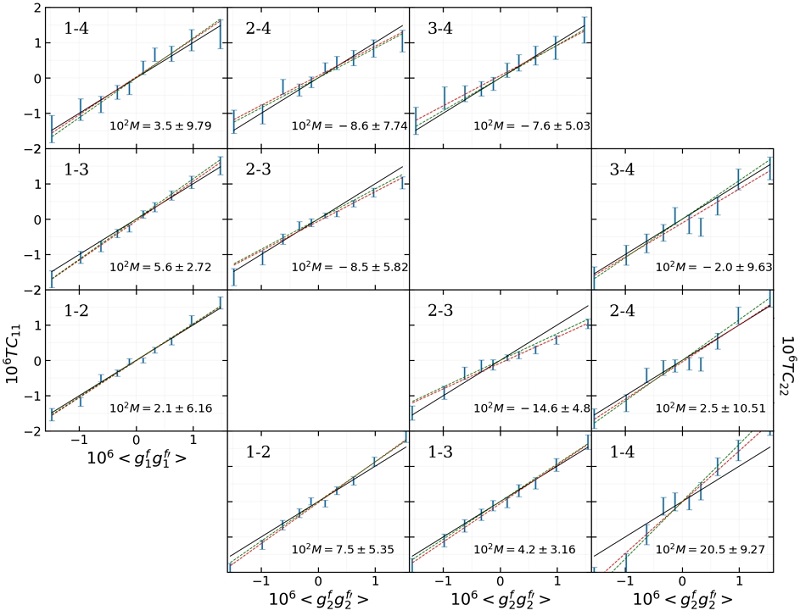}
    \caption{TC between different redshift bins as a consistency check. In each panel, the label on the upper-left corner shows the two bin indices involved in the measurement. The multiplicative bias $M$ is derived by fitting $y=(1+M)^2x$ to the data points. The best-fit line is shown in red in each panel. $M$ is expected to be consistent with the biases $m^{i}$ ("i" stands for the index of the redshift bin) estimated from the auto-correlations, i.e., we expect:$(1+m^{i})(1+m^{j})=(1+M)^2$. For this purpose, we also show in each panel the green line defined as $y=(1+m^i)(1+m^j)x$, using the bias values from table.\ref{tab:mc_of_db}. In this test, we only use the galaxy pairs with separation larger than 170 degrees.}
    \label{fig:cons} 
\end{figure*}

\begin{figure}
\centering
\includegraphics[scale=0.7]{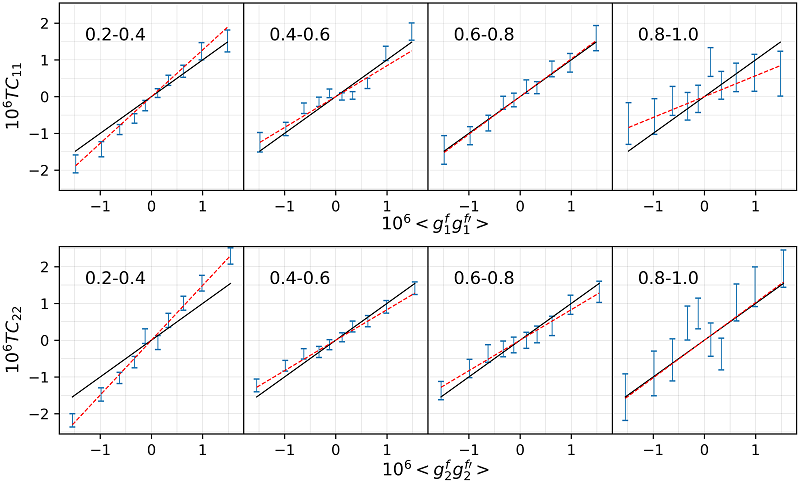}
\caption{The TC of galaxies in different redshift bins. The photo-z catalog is from \cite{ref:Zou2019}. The definitions of the red dashed lines and black solid lines are same as those in fig.\ref{fig:2Dfd_db}.}
\label{fig:2Dfd_db_zh}
\end{figure}

For the tomography study, we use TC to calibrate the shear biases in each redshift bin. We divide the redshift range of $[0.2, 1.0]$ evenly into 4 bins. The TC results of these bins are shown in fig.\ref{fig:2Dfd_db}. The definitions of the axes and the meanings of the black and red lines in the plots are the same as those in fig.\ref{fig:corr_fd}. To our surprise, there is a significant multiplicative biases for both $m_1$ and $m_2$ in a number of redshift bins. The values of the biases are given in table.\ref{tab:mc_of_db}. As a consistency check, we perform the original FD tests (shear stacking) for the galaxies in these redshift bins. The results are shown in fig.\ref{fig:1Dfd_db} and table.\ref{tab:mc_of_1d}. The biases are consistent with those from the TC in all the bins. 

As a further consistency test, we also measure the TC between two different redshift bins. The results are shown in fig.\ref{fig:cons}. The label on the upper-left corner of each panel show the indices of the two redshift bins. The bin index of 1, 2, 3, 4 refers to the redshift range of $[0.2, 0.4]$, $[0.4, 0.6]$, $[0.6, 0.8]$, $[0.8, 1]$ respectively. The definitions of the axes as well as the black solid line and the red dashed line are the same as those in fig.\ref{fig:2Dfd_db}. In this calculation, we only use the galaxy pairs separated by more than 170 degrees. Note that from the auto-correlation of each redshift bin we have already obtained its multiplicative bias $m^{i}$ with i=1, 2, 3, 4 (note that the upper index on $m$ refers to the redshift bin index). We expect that the joint multiplicative bias $\mathcal{M}$ of the TC between two distinct bins is given by $(1+\mathcal{M})^2=(1+m^{i})(1+m^{j})$. These predictions are shown in fig.\ref{fig:cons} with the green lines given by $y=(1+\mathcal{M})^2x$. One can see that the red and green lines in each plot agree well with each other, implying that the multiplicative biases derived from the TC test are reliable. 

It is still unclear to us why the redshift binning causes the multiplicative biases, which can be as high as 10\% or even more. There seem to be a strong selection effect associated with the photo-z. If we switch to a different photo-z catalog from \cite{ref:Zou2019}, the biases persist, as shown in fig.\ref{fig:2Dfd_db_zh} and table.\ref{tab:mc_of_db_zh}. Their amplitudes are similar to the original results in table.\ref{tab:mc_of_db}. 

We give more discussions in the \S\ref{sec:conclu_disc} regarding the bias. Interestingly, in the DES Y3 shear calibration simulation \cite{ref:MacCrann2022}, they also find a redshift-dependent multiplicative bias shown in different bins. We note that unlike the conclusion of \cite{ref:MacCrann2022}, the bias found in this work is NOT due to blending in our test, because the blended sources experience the same FD effect. Given that the FQ shear measurement does not need assumptions about the regularity of the galaxy morphology, we do not expect shear bias to rise in the FD test. 

In the next section, we implement these biases in our tomography study, and check their influence on the cosmological constraints.

\begin{table*}
\centering
\begin{tabular}{ccccc}
\hline
    z &$10^2\times m_1$   & $10^8\times \mathcal{C}_1 $  &$10^2\times m_2$  & $10^8\times \mathcal{C}_2$ \\
\hline
    [0.1-$\infty$] &  -2.0$\pm$1.72& 4.1$\pm$2.6& 0.5$\pm$0.78 &6.4$\pm$1.53  \\
\hline
[0.2-0.4]	 & 17.5 $\pm$ 3.02	 & -5.2$\pm$3.71 &
           	   25.6 $\pm$ 3.26	 & -3.6$\pm$5.26 \\ 
\hline
[0.4-0.6]	 & -11.8 $\pm$ 5.34	 & 7.0$\pm$6.69 &
           	   -10.9 $\pm$ 1.61	 & 2.1$\pm$2.54 \\ 
\hline
[0.6-0.8]	 & -3.2 $\pm$ 3.84	 & -14.4$\pm$4.70 &
           	   -15.4 $\pm$ 2.70	 & -3.3$\pm$4.04 \\ 
\hline
[0.8-1.0]	 & -4.3 $\pm$ 8.72	 & -8.5$\pm$11.30 &
           	   12.9 $\pm$ 5.51	 & 25.6$\pm$9.45 \\ 
\hline
\end{tabular}
\caption{The multiplicative and additive biases for different redshift bins from TC. The photo-z catalog is from \cite{ref:zhou2021}.}
\label{tab:mc_of_db}
\end{table*}

\begin{table*}
\centering
\begin{tabular}{ccccc}
\hline
    z &$10^2\times m_1$ & $10^4\times c_1$   &$10^2\times m_2$ & $10^4\times c_2$ \\
\hline
    [0.1-$\infty$] &  3.2$\pm$4.36& -0.52$\pm$0.35& 0.9$\pm$4.18 &-0.44$\pm$0.35  \\
\hline
[0.2-0.4]	 & 16.3 $\pm$ 3.64	 & -0.8$\pm$0.29 &
           	   24.9 $\pm$ 8.28	 & -1.7$\pm$0.70 \\ 
\hline
[0.4-0.6]	 & -6.6 $\pm$ 9.25	 & -0.2$\pm$0.74 &
           	   -13.3 $\pm$ 3.57	 & 0.6$\pm$0.30 \\ 
\hline
[0.6-0.8]	 & -7.2 $\pm$ 6.71	 & -0.7$\pm$0.54 &
           	   -14.7 $\pm$ 8.24	 & 1.8$\pm$0.70 \\ 
\hline
[0.8-1.0]	 & -9.7 $\pm$ 17.97	 & 1.6$\pm$1.44 &
           	   11.1 $\pm$ 26.44	 & -4.1$\pm$2.24 \\ 

\hline
\end{tabular}
\caption{The multiplicative and additive biases for different redshift bins from the original FD test (shear stacking). The photo-z catalog is from \cite{ref:zhou2021}.}
\label{tab:mc_of_1d}
\end{table*}

\begin{table*}
\centering
\begin{tabular}{ccccc}
\hline
    z &$10^2\times m_1$   & $10^8\times \mathcal{C}_1$  &$10^2\times m_2$  & $10^8\times \mathcal{C}_2$ \\
\hline
[0.2-0.4]	 & 13.6 $\pm$ 2.44	 & -9.0$\pm$3.02 &
           	   24.6 $\pm$ 2.90	 & 0.1$\pm$4.87 \\ 
\hline
[0.4-0.6]	 & -8.0 $\pm$ 6.11	 & 1.7$\pm$7.27 &
           	   -8.5 $\pm$ 1.59	 & 4.9$\pm$2.53 \\ 
\hline
[0.6-0.8]	 & 1.3 $\pm$ 2.93	 & 3.2$\pm$3.74 &
           	   -8.5 $\pm$ 2.16	 & -0.8$\pm$3.49 \\ 
\hline
[0.8-1.0]	 & -21.6 $\pm$ 7.13	 & 16.2$\pm$10.05 &
           	   1.3 $\pm$ 11.01	 & 23.6$\pm$17.44 \\ 
\hline
\end{tabular}
\caption{The multiplicative and additive biases for the different redshift bins from TC, using the photo-z catalog from \cite{ref:Zou2019}.}
\label{tab:mc_of_db_zh}
\end{table*}

\section{Impact on Cosmological Constraints}
\label{cosmo}

\subsection{Theory}\label{sec:tomo}

The conventional shear-shear correlations are measured using the tangential and cross shear components ($\gamma_t$ and $\gamma_\times$), which are defined as $\gamma_t+i \gamma_{\times}=- (\gamma_1+i \gamma_2 ) e^{-2 \alpha i},\alpha$ is the angle between the x-axis and the line connecting the galaxy pair. Correspondingly, the correlations are defined as:
\begin{equation}
    \begin{aligned}
    \xi_{\pm}(z_1,z_2,\Delta\vec{\theta})&=<\gamma_t(z_1,\vec{\theta})\gamma_t(z_2,\vec{\theta}+\Delta{\vec{\theta}})>\\
    &\pm<\gamma_{\times}(z_1,\vec{\theta})\gamma_{\times}(z_2,\vec{\theta}+\Delta{\vec{\theta}})>    
    \end{aligned}
\end{equation}
In this paper, similar to the work of \cite{ref:zjliu2023}, we also consider another way: correlations with only $\gamma_1$ or $\gamma_2$, i.e., $\xi_{ii} =<\gamma_i(z_1,\vec{\theta})\gamma_i(z_2,\vec{\theta}+\Delta{\vec{\theta}})>$, which i =1 or 2. Theoretically, we expect $\xi_{11}=\xi_{22}=\xi_{+}/2$. We use these different types of correlations to constrain cosmology. 

In the theoretical model of $\xi_{+/-}$, one needs to consider the intrinsic alignment. For our purpose, we simply adopt the recipe of \cite{ref:heymans2013}, in which the shear-shear correlation can be written as:
\begin{equation}
    \xi_{\pm}=\xi_{GG\pm}+\xi_{GI\pm}+\xi_{II\pm}
\end{equation}
where the subindex ”G” represents the cosmic shear, and ”I” stands for the intrinsic galaxy shape. $\xi_{+/-}$ for galaxy pairs with the angular distance $\theta$ are given by:
\begin{equation}
\begin{aligned}
&\xi_{ \pm}^{i j}(\theta)=\int_0^{\infty} \frac{d \ell}{2 \pi} \ell C^{i j}(\ell) J_v(\ell \theta),\\
&C^{i j}(\ell)=\int_0^{\infty} d z \frac{W^i(z) W^j(z)}{\chi(z)^2} P_\delta (\frac{\ell}{\chi(z)}, z )
\label{eq:xi_ij}
\end{aligned}
\end{equation} 
where $J_{\nu}$ is the first kind Bessel function, and $\nu$ is 0/4 for $\xi_{+}$/$\xi_{-}$. $\chi(z)$ is the comoving radial distance at z. $W^i(z)$ represents the kernel including the contributions from both lensing and the intrinsic alignment, i.e., $W^i(z)=W_{\mathrm{G}}^i(z)+W_I^i(z)$. The kernel for shear is given by:
\begin{equation}  
W_{\mathrm{G}}^i(z)=\frac{3}{2} \Omega_m \frac{H_0^2}{c^2} \frac{\chi(z)}{a(z)} \int_z^{\infty} d z^{\prime} n^i (z^{\prime} ) \frac{\chi (z^{\prime} )-\chi(z)}{\chi (z^{\prime} )}, 
\label{equ:xi_pm_c_ij}
\end{equation}
in which $n^i(z)$ is the normalized redshift distributions, $a(z)$ is the scale factor, $H_0$ is the Hubble constant, and c is
the speed of light. For simplicity we assume a flat universe in this paper. For the IA kernel we
adopt the Nonlinear Alignment Model (NLA) \cite{ref:bridle2007}:
\begin{equation}
        W_I^i(z)=-\frac{A_{IA}C_1\rho_c\Omega_m}{D(z)}n^i(z)
\end{equation}
where $A_{IA}$ is a free parameter to show the amplitude of IA. $\rho_c$ is the
critical density, D(z) is the normalized growth factor. The normalization constant $C_1$ is adjustable and can be set to $5\times10^{-14}h^{-2}\mathrm{M}_{\odot}^{-1}\mathrm{Mpc}^3$ to align with the observational findings\citep{ref:brown2002}. In this case, the fiducial value of $A_{IA}$ is 1. 

$P_\delta$ in eq.(\ref{eq:xi_ij}) is the nonlinear matter power spectrum, which could be strongly affected by the baryonic effect. In our work, we use the baryonic correction model (BCM, \cite{ref:bcm_schneider2015}) which parameterizes the influence of gas and stars on the total matter density. There are two crucial parameters of the BCM model: the mass fraction of ejected gas ($M_c$) and the ejection radius (which depends on the parameter $\eta_b$). We choose their fiducial values to be: $M_c=1.2M_{\odot}/h, \eta_b=0.5$. These values are consistent with simulations and observations \citep{ref:bcm_schneider2015}. Overall, our theoretical predictions are calculated by the Core Cosmology Library (CCL, \cite{ref:chisari2019}), in which the nonlinear evolution is described by the halofit model \citep{ref:Smith2003, ref:Takahashi2012}.

\subsection{Redshift Distribution}\label{sec:redshift}

\begin{figure}
    \centering
    \includegraphics[scale = 0.29]{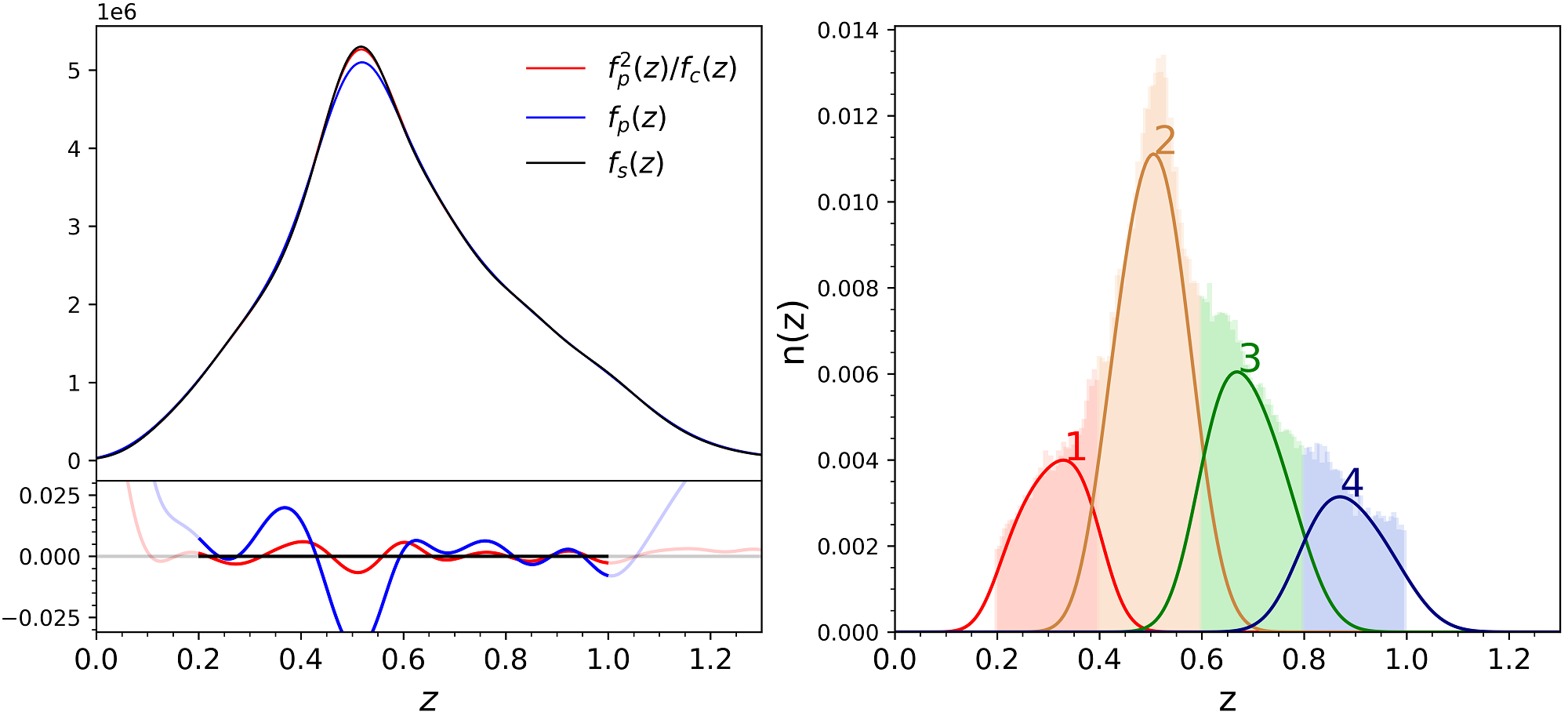}
    \caption{The left panel shows how well the true redshift distribution $f_s(z)$ can be recovered with $f_p^2(z)/f_c(z)$ using the method described in \S\ref{sec:redshift}. The y-axis of its upper panel has an arbitrary unit, and that of the lower panel shows the relative differences between $f_p(z)$ \& $f_s(z)$ (in blue), and $f_p^2(z)/f_c(z)$ \& $f_s(z)$ (in red). The right panel shows the redshift distributions of the four photo-z bins used for our tomography study.}
    \label{fig:bins}
\end{figure}

In our tomography study, we use the photo-z catalog from \cite{ref:zhou2021}. For each redshift bin selected according to the photometric redshifts, we hope to get its true redshift distribution $n^i(z)$ used in eq.(\ref{eq:xi_ij}). It is related to the photo-z distribution $f_p(z_p)$ through the following equation:
\begin{equation}
n^i(z_s)=\int_{z^i_{min}}^{z^i_{max}}P(z_s|z_p)f_p(z_p)dz_p
\end{equation}
in which $z^i_{min}$ and $z^i_{max}$ are the lower and upper bounds of the bin, $f_p(z_p)$ is the overall photo-z distribution, and $P(z_s|z_p)$ is the probability that a galaxy of $z_p$ has a true redshift of $z_s$. Unfortunately, $P(z_s|z_p)$ is not directly known to us. Its counterpart, $P(z_p|z_s)$, is usually more directly accessible by studying the performance of photo-z reconstruction using simulations. They are related through the Bayes theorem: 
\begin{equation}
    P(z_s|z_p)f_p(z_p) = P(z_p|z_s)f_s(z_s), 
\label{eq:bys}
\end{equation}
where $f_s(z_s)$ is the distribution of the true redshift, which is neither known. If we can get an approximate form of $f_s(z_s)$, the form of $P(z_s|z_p)$ can then be derived. For this purpose, we assume $P(z_p|z_s)$ is a Gaussian function with a pre-determined scatter $\sigma_z$. 
Integrating both sides of eq.(\ref{eq:bys}) over all possible values of $z_s$, we get:
\begin{equation}
    f_p(z_p) = \int f_s(z_s)P(z_p|z_s)dz_s
\label{equ:gzp}
\end{equation}
In principle, $f_s(z_s)$ can be derived by inverting the convolution in eq.(\ref{equ:gzp}). In practice, we find that the solution from such an inversion is not very stable. Assuming $\sigma_z$ is small, the integration in eq.(\ref{equ:gzp}) can be well approximated as:
\begin{equation}
    f_p(z_p) \approx f_s(z_p)\int P(z_p|z_s)dz_s=f_s(z_p)*g(z_p)
\end{equation}
To get the form of $g(z)$, one can further convolve $f_p(z)$ with the same Gaussian kernel $P(z_p|z_s)$ to get $f_c(z)$, and therefore $g(z_p)\approx f_c(z)/f_p(z)$. Consequently, we get $f_s(z)\approx f_p(z)/g(z)\approx f_p^2(z)/f_c(z)$. In the left panel of fig.\ref{fig:bins}, we show using simulations how well the $f_s(z)$ derived using the above method can recover the true redshift distribution. The right panel of the same figure shows the recovered redshift distribution for the four photo-z bins. We have assumed that $P(z_p|z_s)$ is a Gaussian kernel with $\sigma_z= 0.03*(1+z)$. This is a reasonable choice according to both \cite{ref:zhou2021} and our another work (Li et al., in preparation). This assumption is used in our tomography study in the next section. We have checked that our main results do not change much if we assign $\sigma_z= 0.05*(1+z)$ instead.


\subsection{Tomography}

\begin{figure*}
    \centering
    \includegraphics[scale = 0.7]{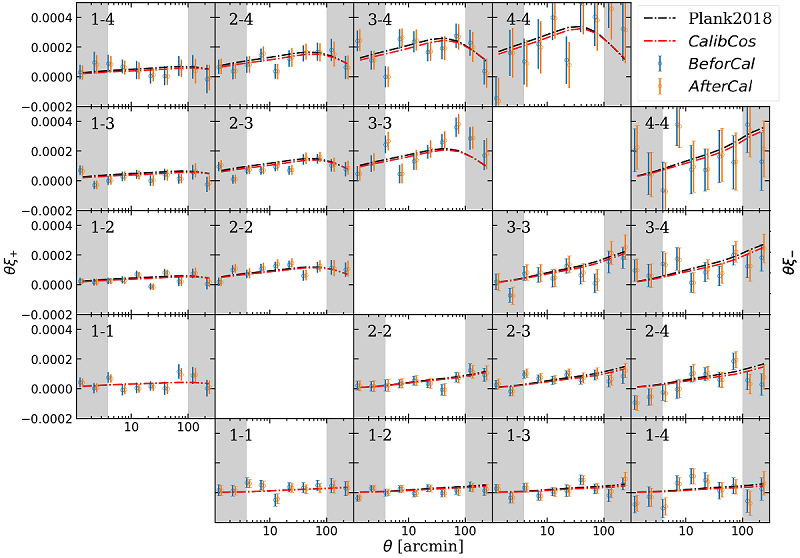}
    \caption{The tomographic shear-shear correlation functions (2PCF). The indices of the redshift bins are shown in the upper-left corner of each panel. The blue and orange data points show the 2PCF without and with the corrections from TC respectively. The blue points are shifted to their left by 10\% for the clearness of view. The black theoretical curves show the prdictions from \cite{ref:planck_result} with $A_{IA}=1$. The red curves are from the best-fit cosmological parameters of the orange data points. The data in the shaded areas are excluded in our cosmological constraints. }
    \label{fig:tomo}
\end{figure*}


Fig.\ref{fig:tomo} shows the measured two-point correlation functions $\xi^{ij}_{\pm}(\theta)$ for all the pairs of redshift bins.
The angular range is from 1 arcmin to 300 arcmin.
For the cosmological constraint, we do not use the data points 
below 4 arcmin or above 100 arcmin. The blue points are the results of $\xi_{\pm}$ without any calibration. The orange data points are the results of $\xi_{\pm}$ after we incorporate the corrections from TC, i.e., we multiply the shear estimator $G$ of each galaxy by $(1+m)^{-1}$ from the TC according to its redshift bin. 
For clarity in the figure, we move the blue points slightly to the left of their original places. The error bars and the covariance matrix of the data points are estimated using the Jackknife method. We use the K-Means clustering method \citep{ref:Pedregosa2011} to divide the galaxies into 200 groups.
The red dashed line is the best-fit cosmological prediction to the orange data points (after calibration). The black dashed lines are calculated from the cosmological parameters of PLANCK \citep{ref:planck_result}, with $A_{IA}=1$.

\begin{figure}
    \centering
    \includegraphics[scale = 1.]{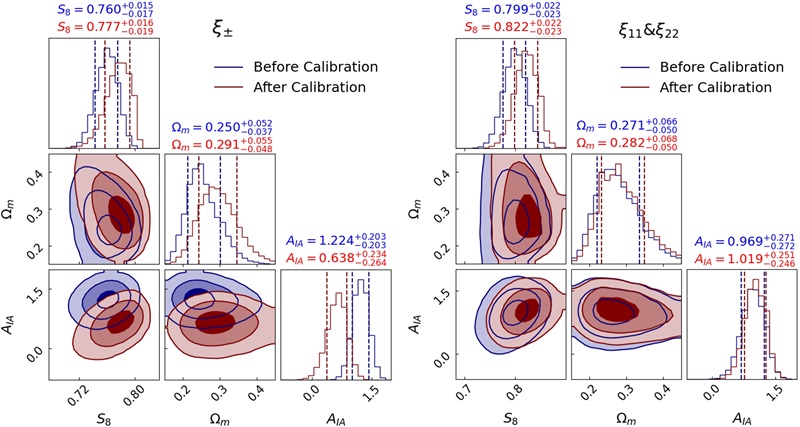}
    \caption{Constraints on $S_8$, $\Omega_m$, and $A_{IA}$ using either $\xi_+$ \& $\xi_-$ or $\xi_{11}$ \& $\xi_{22}$. The error contours in each panel and each color correspond to the 68\%, 95\% and 99\% confidence level respectively. The blue and red contours show the constraints without and with the corrections from TC respectively.}
    \label{fig:xipm_wo}
\end{figure}

For the cosmological constraint, we use the standard Markov-Chain Monte Carlo (MCMC) method with EMCEE\cite{ref:emcee} for cosmological parameter estimation. We treat $A_{IA}$ of the intrinsic alignment as a free parameter. 
Our main results are shown in fig.\ref{fig:xipm_wo}. The blue contours and the red ones are for the best-fit cosmological parameters from the blue and orange data points in fig.\ref{fig:tomo} respectively. Without correcting for the redshift-dependent bias, we get: $S_8 = 0.760^{+0.015}_{-0.017},\Omega_m = 0.250^{+0.052}_{-0.037}, A_{IA} = 1.224^{+0.203}_{-0.203}$. After we incorporate the corrections of the biases estimated from TC in each redshift bin, we get: $S_8=0.777^{+0.016}_{-0.019},\Omega_m=0.291^{+0.055}_{-0.048}, A_{IA}=0.638^{+0.234}_{-0.264}$. It is interesting and perhaps important to note that the $S_8$ value has about 1$\sigma$ increase as a result of the calibration from TC. 

\begin{figure*}
    \centering
    \includegraphics[scale = 0.7]{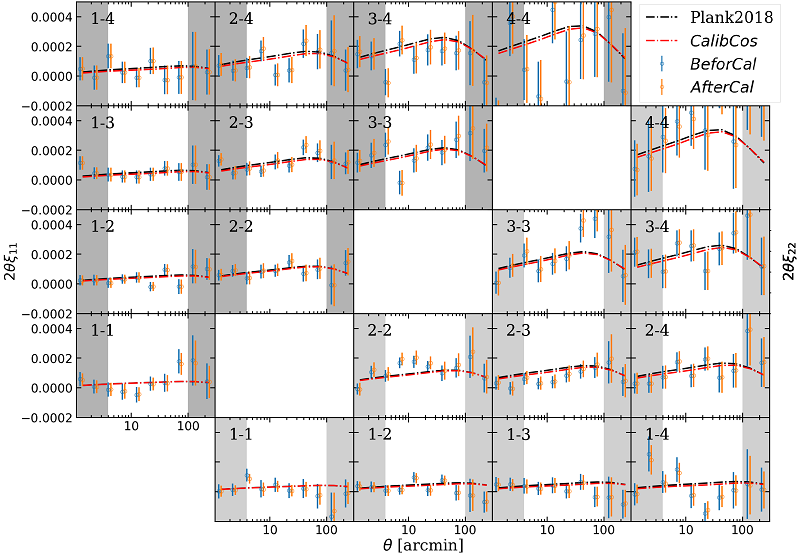}
    \caption{Similar to fig.\ref{fig:tomo}, except that the cosmological constraints are from $\xi_{11}$ \& $\xi_{22}$.}
    \label{fig:g1g1}
\end{figure*}

\begin{figure}
    \centering
    \includegraphics[scale = 0.3]{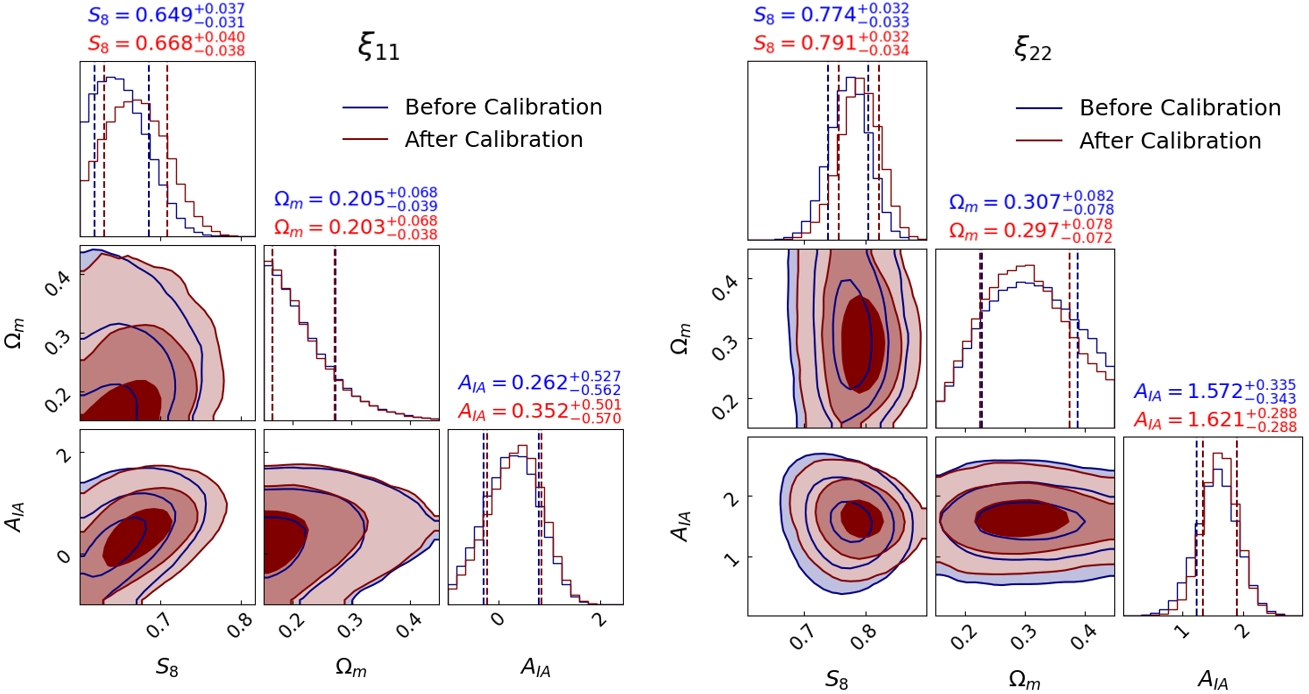}
    \caption{Similar to fig.\ref{fig:xipm_wo}, but for the constraints from $\xi_{11}$ (left) and $\xi_{22}$ (right) separately.}
    \label{fig:12t_wo}
\end{figure}

For a comparison, we show the results of $2*\theta\xi_{11}$ and $2*\theta\xi_{22}$ in fig.\ref{fig:g1g1}. 
The definitions of the blue and orange data points, as well as the black and red lines are the same as those in fig.\ref{fig:tomo}. The cosmological constraint using $\xi_{11}$ and $\xi_{22}$ together is shown in the right panel of fig.\ref{fig:xipm_wo}. Similarly, we find that the calibration from TC leads to a $1\sigma$ increase in $S_8$. It is worth noting that the data quality of $\xi_{11}$ and $\xi_{22}$ is somewhat worse than that of $\xi_+$ and $\xi_-$. The cosmological constraints from $\xi_{11}$ or $\xi_{22}$ separately are shown in fig.\ref{fig:12t_wo}, from which one can see that the quality of $\xi_{11}$ is even worse than $\xi_{22}$. Similar phenomenon has been reported in our another work \citep{ref:zjliu2023}.

In the DECaLS shear catalog, the first and second shear components are defined in the local coordinates along the directions of "RA" and "Dec", and the CCDs are always lined up with the same direction in the survey, the image quality therefore naturally inherits certain anisotropy from hardware imperfectness. This problem is mitigated by rotating the shear estimators in the measurement of $\xi_{+/-}$, yielding results of seemingly higher qualities. However, we shall instead take this as a caution for the existence of unresolved systematic issues in image processing. 

\begin{figure*}
    \centering
    \includegraphics[scale = 0.5]{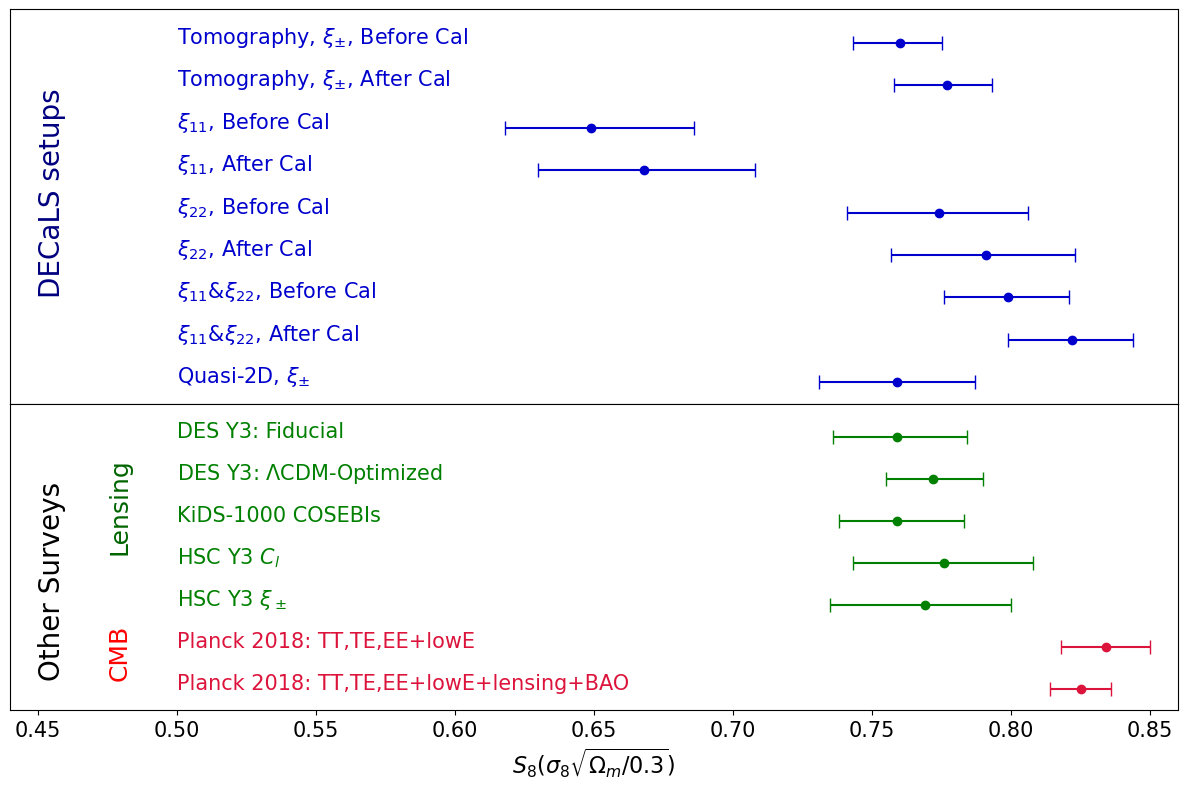}
    \caption{Summary of the $S_8$ constraints from various consistency tests, and from other lensing surveys and Planck.}
    \label{fig:s8_summary}
\end{figure*}

\begin{figure}
    \centering
    \includegraphics[scale = 0.7]{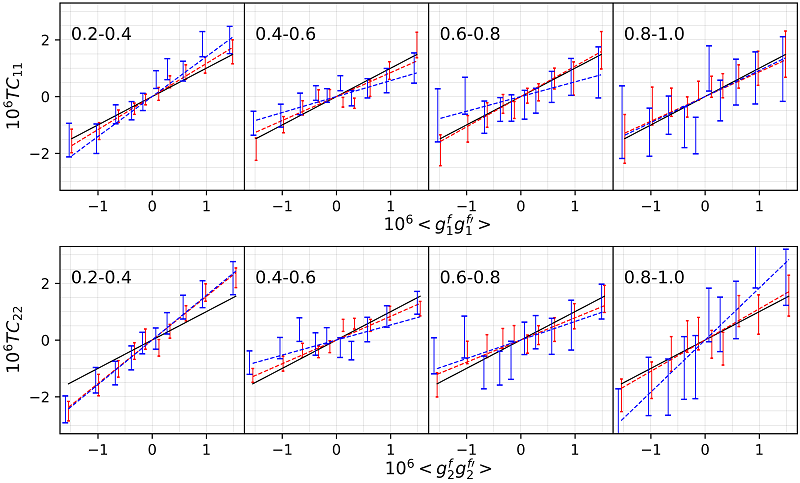}
    \caption{Comparison of the TC results from galaxies of different PSF sizes. In each panel, the red and blue points are from galaxies with smaller (${\mathrm{FWHM_{PSF}}} < 1.4"$) and larger (${\mathrm{FWHM_{PSF}}} > 1.4"$) PSF sizes respectively. The dashed lines are the best-fit of the data points. The redshift range of the galaxies is indicated in the upper-left corner of each panel.}
    \label{fig:diff_psf}
\end{figure}

\begin{figure}
    \centering
    \includegraphics[scale = 0.7]{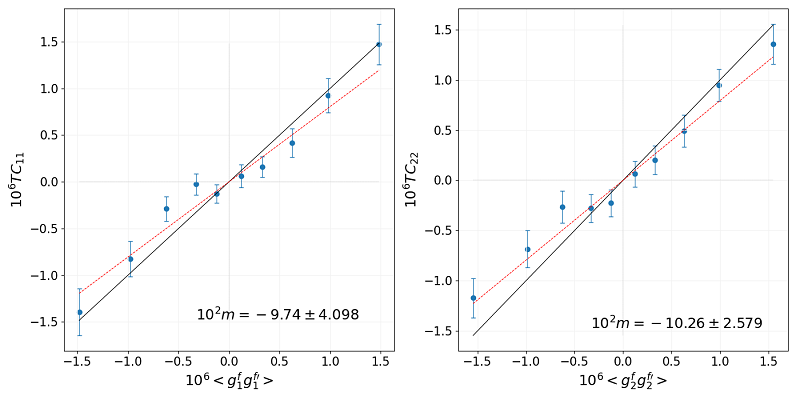}
    \caption{TC of galaxies in the redshift range of [0.4, $\infty$]. The black solid lines and the dashed lines have the same meanings as those in fig.\ref{fig:corr_fd}.}
    \label{fig:tc_0.4_10}
\end{figure}

\section{Conclusion and Discussions}
\label{sec:conclu_disc}

Tele-correlation (TC), the correlation of the shear estimators of two galaxies separated by a large distance ($\gtrsim$ 100 degree), can be used to calibrate the multiplicative and additive biases in shear-shear correlations. In TC, the correlation signal comes from the field distortion (FD), which can be retained in the shear catalog if the shear estimators are associated with individual exposures. 
We demonstrate this idea with the DECaLS shear catalog produced by the Fourier\_Quad pipeline. 
With all the distant galaxy pairs ( $>140$ degrees), we do not observe any bias in the correlation signal at the level of $10^{-6}$.
However, to our surprise, significant multiplicative biases can arise if the TC test is performed in individual photo-z bins. The reason is still not known to us.

Using the conventional tomographic shear-shear correlations $\xi_{+/-}$, we try to place constraints on the cosmological parameters, and meanwhile study the impact of the biases (from TC) on the results. We use the galaxies in the redshift range of [0.2, 1.0], and the angular range from 4 to 100 arcmins for calculating the correlations. We choose the angular range conservatively to avoid the effects of some unknown physics at very small and large scales. In our cosmological model, we adopt the NLA model for the intrinsic alignment, and BCM for the impact of baryons on the density power spectrum. Without the correction for the biases, we obtain $S_8 = 0.760^{+0.015}_{-0.017}$, $\Omega_m = 0.250^{+0.052}_{-0.037}$ and $A_{IA} = 1.224\pm0.203$. After incorporating the multiplicative biases due to redshift binning, the cosmological constraints become: $S_8=0.777^{+0.016}_{-0.019},\Omega_m=0.291^{+0.055}_{-0.048}, A_{IA}=0.638^{+0.234}_{-0.264}$. There is about a $1\sigma$ increase in the best-fit value of $S_8$. 


In fig.\ref{fig:s8_summary}, we summarize our constraints on $S_8$ from different types of shear-shear correlations, including $\xi_{ii}$ ($=<\gamma_i\gamma_i>$, i=1 or 2), separately or jointly, with or without calibration from TC. In these cases, we see a large variation of the constraints, particularly those with $\xi_{11}$. This is an indication of the potential systematic biases in the shear catalog, likely due to the image quality, as discussed in \citep{ref:zjliu2023}. We also present the $S_8$ constraints from the other lensing surveys such as the fiducial $\Lambda$CDM-optimized analyses in DES Y3 \citep{ref:des_y3_shear}, KiDS-1000 cosmology \citep{ref:asgari2021}, the analyses of $C_l$ and $\xi_{\pm}$ in HSC Y3 \citep{ref:Lixc2023}. The Planck results \citep{ref:planck_result} with the baseline TT, TE, EE+LowE are also included. Our tomographic lensing constraints on $S_8$ are mostly consistent with those of the other lensing surveys, having about $2-3\sigma$ tension with the Planck results.

Unfortunately, we do not yet have a good understanding of the significant multiplicative biases found in our TC test. It is likely a selection effect because the galaxy redshift affects multiple image properties, including the apparent magnitude, size, shape (after convolving with the PSF). 
The origin of these biases remains elusive, and we will investigate the underlying mechanisms in a future work. For now, to go a little further in this topic, we divide the galaxy sample into two groups based on the PSF size. We measure the tele-correlations for the first and second shear components ($TC_{11}$ and $TC_{22}$), and show the results in fig.\ref{fig:diff_psf}. The black solid line in each panel is the 'y=x' line. The red lines are from the galaxies with smaller PSF size (${\mathrm{FWHM_{PSF}}} < 1.4"$), and the blue ones for larger PSF (${\mathrm{FWHM_{PSF}}} > 1.4"$). It is interesting to note that the multiplicative biases seem to decrease with the PSF size. This fact agrees with our intuition: shape measurement is more sensitive to the galaxy size/redshift when the spatial resolution of the image is poorer, i.e., when the PSF size is larger, thereby leading to a larger selection bias. This is a good news for those surveys with small PSFs, such as HSC (Liu et al., in preparation).

Some previous works (e.g., \cite{ref:jqwang2022,ref:mfong2022,Xu_2023}) have used the FQ shear catalog of DECaLS to study a number of statistics mainly in galaxy-galaxy lensing. We realize that in those studies, since they only use the background galaxies with redshifts larger than certain threshold (e.g., the lens redshift plus some given value), the selection in redshift necessarily cause some bias that needs to be corrected for. For example, we test the TC of the background sample within the redshift range of $[0.4,\infty]$. The results are presented in fig.\ref{fig:tc_0.4_10}, which show a negative multiplicative bias of about $10\%$. Note that this result can also be obtained with the original FD test. The actual correction should be measured more carefully, as galaxies of different redshifts are usually weighted by the critical surface density in galaxy-galaxy lensing. The details will be discussed in our future work. 

On the other hand, it is encouraging to note that the FD test offers us a  convenient onsite calibration of the shear biases, no matter whether the bias is caused by PSF inaccuracy, instrumental effects, or selection effects. To our knowledge, the only case in which the FD test may not work is about the shear bias due to image blending, i.e., galaxy pairs that are close in angular space, but far in redshift space. Since in this case, the two galaxies cannot be treated as a single object in shear measurement, and the FD cannot tell apart the two objects. 


\acknowledgments
This work is supported by
the National Key Basic Research and Development Program of
China (2023YFA1607800, 2023YFA1607802), the NSFC
grants (11621303, 11890691, 12073017), and the science research grants from China Manned Space Project (No.CMS-CSST-2021-A01). The computations in this paper were run on the $\textit{Siyuan}$ cluster supported by the Center for High-Performance Computing at Shanghai Jiao Tong University.

\bibliography{main.bib}
\bibliographystyle{JHEP}

\end{document}